# Freeze cast porous barium titanate for enhanced piezoelectric energy harvesting

J. I. Roscow[a,*], Y. Zhang[a,*], M. J. Krasny[a], R. W. C. Lewis[b], J. Taylor[c], & C. R. Bowen[a]

[a] Department of Mechanical Engineering, University of Bath, Bath, UK
[b] Renishaw Plc., Wooton-under-Edge, Gloucestershire, UK
[c] Department of Electrical and Electronic Engineering, University of Bath, Bath, UK

**Abstract**

Energy harvesting is an important developing technology for a new generation of self-powered sensor networks. This paper demonstrates the significant improvement in the piezoelectric energy harvesting performance of barium titanate by forming highly aligned porosity using freeze casting. Firstly, a finite element model demonstrating the effect of pore morphology and angle with respect to poling field on the poling behaviour of porous ferroelectrics was developed. A second model was then developed to understand the influence of microstructure-property relationships on the poling behaviour of porous freeze cast ferroelectric materials and their resultant piezoelectric and energy harvesting properties. To compare with model predictions, porous barium titanate was fabricated using freeze casting to form highly aligned microstructures with excellent longitudinal piezoelectric strain coefficients, $d_{33}$. The freeze cast barium titanate with 45 vol.% porosity had a $d_{33}$ = 134.5 pC/N, which compared favourably to $d_{33}$= 144.5 pC/N for dense barium titanate. The $d_{33}$ coefficients of the freeze cast materials were also higher than materials with uniformly distributed spherical porosity due to improved poling of the aligned microstructures, as predicted by the models. Both model and experimental data indicated that introducing porosity provides a large reduction in the permittivity ($\varepsilon_{33}^\sigma$) of barium titanate, which leads to a substantial increase in energy harvesting figure of merit, $d_{33}^2/\varepsilon_{33}^\sigma$, with a maximum of 3.79 pm$^2$/N for barium titanate with 45 vol.% porosity, compared to only 1.40 pm$^2$/N for dense barium titanate. Dense and porous barium titanate materials were then used to harvest energy from a mechanical excitation by rectification and storage of the piezoelectric charge on a capacitor. The porous barium titanate charged the capacitor to a voltage of 234 mV compared to 96 mV for the dense material, indicating a 2.4-fold increase that was similar to that predicted by the energy harvesting figures of merit.

1. Introduction

Energy harvesting, the process of recapturing energy from ambient sources, such as mechanical vibrations and waste heat, and converting it to useful electrical energy, has received increasing attention in recent years with the development of low powered electronics and wireless sensor technologies [1]. Ferroelectric ceramics are of particular interest in this field due to their ability to directly convert vibrational energy to electrical energy via the piezoelectric effect and thermal fluctuations into electrical energy via the pyroelectric effect [2, 3]. Figures of merit for both piezoelectric and pyroelectric energy harvesting have been derived based on the change of polarisation of a poled ferroelectric material due to an applied stress (piezoelectric) or change in temperature (pyroelectric). The piezoelectric harvesting figure of merit for off-resonance, low frequency (<<100 kHz) vibration is given by [4]:

$$FoM_{ij} = \frac{d_{ij}^2}{\varepsilon_{33}^\sigma} \quad (1)$$

where $d_{ij}$ is the piezoelectric strain coefficient (the subscripts denote the direction of applied stress ($j$) with respect to the poling direction ($i$)) and $\varepsilon_{33}^\sigma$ is the permittivity at constant stress. The pyroelectric energy harvesting figure of merit is given by [5]:

$$F_E' = \frac{p^2}{c_E^2 \cdot \varepsilon_{33}^\sigma} \quad (2)$$

where $p$ is the pyroelectric coefficient and $c_E$ is the volume specific heat capacity. A simple analysis of the figures of merit above demonstrate that when selecting a ferroelectric to harvest mechanical or thermal vibrations it should have a high piezoelectric strain or pyroelectric coefficient and a low permittivity. However, ferroelectric materials with the highest piezoelectric and pyroelectric coefficients often have high permittivity. One approach to improve the figures of merit is to introduce porosity, which has been shown to provide a cost-effective way of tuning the properties of these materials by reducing the permittivity whilst maintaining a relatively high $d_{33}$ [6] or $p$ [7].

Directly after sintering, a polycrystalline ferroelectric ceramic initially exhibits no piezoelectric or pyroelectric properties as domains within the material are randomly orientated [8]. To make the ferroelectric material electro-active they are poled by applying an electric field to orientate the ferroelectric domains which exist below the Curie temperature, yielding a net polarisation that provides a piezoelectric and pyroelectric response. It has been shown previously that the decrease in piezoelectric response in porous ferroelectrics is largely due to difficulties in poling these materials as the poling field preferentially concentrates in the low permittivity pores, thereby leading to regions of ferroelectric material remaining unpoled after the poling field is removed [9, 10]. Understanding the effect of the porous structure on the poling behaviour of ferroelectrics may therefore aid the design of composites materials with low permittivity and high $d_{33}$ coefficients. The processing method used to form a porous ceramic structure determines the connectivity, distribution and alignment of the final porous structure [11], and this paper aims to demonstrate the types of microstructure that are necessary to improve piezoelectric energy harvesting capabilities of porous ferroelectric ceramics.

The introduction of porosity into ferroelectric ceramics such as lead zirconate titanate (PZT) and barium titanate (BaTiO$_3$) reduces the permittivity of the ceramic-air composite [12] and has been shown to be beneficial in terms of increasing figures of merit for a number of applications including piezoelectric hydrostatic acoustic sensors [13-17], pyroelectric thermal detection devices [18, 19] and, more recently, for both piezoelectric [6, 10, 20] and pyroelectric energy harvesting [7, 21]. The presence of high levels of uniformly distributed porosity (i.e. isolated 3-0 or interconnected 3-3 pore connectivity, see Fig. 1a and b, respectively) lead to a relatively small decrease in $d_{33}$ up to a porosity fraction of ~50 vol.%, with a large decrease in $d_{33}$ at higher porosity levels[1] [15-17, 20]. Freeze cast PZT-based materials with excellent alignment of both pore and ceramic channels (3-1 and 2-2 connectivity, see Fig. 1c and d, respectively) have been shown to have higher $d_{33}$ coefficients [7, 21-24] than porous PZT with uniformly distributed porosity at similar levels of porosity [15, 16]. For example, a freeze cast PZT with over 65 vol.% porosity was reported to have $d_{33}$ values of up to 91% that of the dense material, whilst the permittivity was found to be just 33% that of dense PZT [22], yielding an exceptionally high piezoelectric energy harvesting figure of merit using Eqn. 1 [6]. Ionotropic gelation has also recently been used to produce highly aligned 3-1 porous PZT with excellent potential for energy harvesting due to their high $d_{33}$ coefficients [25, 26].

---

[1] It should be highlighted that while the $d_{33}$ can remain relatively unchanged, the large reductions in $d_{31}$ of porous ferroelectric materials make them unsuitable for energy harvesters that operate in this mode, i.e. bending mode cantilever devices.

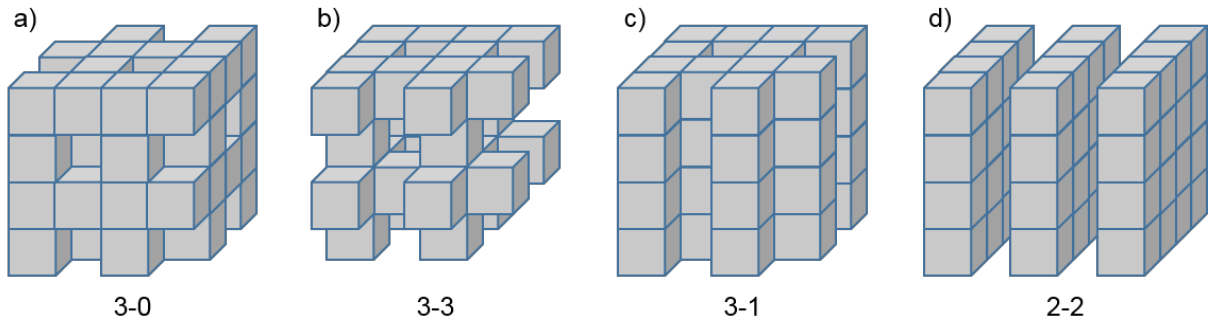

*Figure 1: Schematic demonstrating selected types of connectivity for two-phase composites. The first number refers to the connectivity of the ferroelectric phase and the second number refers to the connectivity of the porosity.*

### 1.1 Effect of porous structure on poling and $d_{33}$

The inhomogeneous field distribution during the poling of porous ferroelectrics is thought to lead to incomplete poling of the system and a decrease in $d_{33}$ compared to the dense material [9, 10]. This occurs as the electric field applied during poling preferentially concentrates in the low permittivity phase, a condition of Gauss' law [27]. There have been several recent investigations into the effect of pore shape, orientation and connectivity on the electric field distribution and polarization-switching dynamics in dielectric materials [28-31]. Porous structures with ideal 3-1 connectivity, i.e. isolated 1D pore channels in a continuous piezoelectric-matrix as in Fig. 1c, have been simulated via finite element analysis and shown to have homogenous electric field distributions throughout the material when an external field is applied parallel to the dielectric phase [28]. In composites with 3-0 (isolated pores, Fig. 1a) and 3-3 (interconnected pores, Fig. 1b) connectivity, an increase in porosity is found to reduce the local electric field in the high permittivity ceramic phase as the electric field preferentially concentrates in the low permittivity pores [29], which partially explains the changes in polarisation-electric field (P-E) loops observed in PZT [7] and $(Pb,Nb)(Ti,Zr)O_3$ (PZTN) [29] with increasing porosity. A detailed study on the effect of pore aspect ratio and orientation with respect to applied electric field demonstrated that elongated pores aligned perpendicular to the applied field exhibited broadening of the P-E loops compared with equi-axed pores, however, the least disruption to electric field was observed when elongated pores were aligned parallel to the electric field [30]. Experimental studies have demonstrated that elongated pores aligned perpendicular to the poling direction led to lower $d_{33}$ values compared to spherical pores [30, 32, 33], which is in agreement with theoretical studies [28-31].

With regards to developing porous ferroelectrics for energy harvesting, microstructures that promote homogenous field distributions throughout the material with local electric fields close to the applied field in both the ceramic phase and the pores are likely to be easier to pole than structures with broad field distributions during the poling process. This would explain the high $d_{33}$ coefficients reported for freeze cast porous piezoelectric materials with an orientated structure [7, 21-24] compared with uniformly distributed porosity [15, 16], which is of benefit for energy harvesting in terms of the relevant figure of merit, see Eqn. 1.

*1.2 Effect of porosity on mechanical properties*

The impact of porosity on mechanical properties are also of importance for applications related to vibrational mechanical harvesting. Porosity increases the mechanical compliance [16, 34] and also decreases the strength of ceramics [35]. However, aligning the high stiffness ceramic phase along the primary loading axis, as is the case of freeze cast porous ceramics, improves the mechanical properties compared to those with uniformly distributed porosity [36]. In order to harvest energy from mechanical loads, porous ferroelectric ceramics must be sufficiently stiff and strong to survive their operating conditions, which is likely to be a further advantage of using highly aligned freeze cast materials rather than those with uniformly distributed porosity.

In this paper both a single pore model and a porous microstructural model are presented to evaluate the effect of the porous structure on the resultant piezoelectric properties and energy harvesting performance. Firstly, the single pore model is used to understand the effect of the angle and orientation of a pore in a barium titanate matrix on the poling behaviour of the material, which provides an explanation of the excellent $d_{33}$ coefficients reported in literature for aligned microstructures. Secondly, the porous microstructural model is used to understand the effect of microstructural features commonly observed in porous freeze cast materials on the piezoelectric and dielectric properties. To provide experimental data, barium titanate with highly aligned porosity has been produced via the freeze casting method to improve the degree of poling and provide an experimental comparison to the observations from the modelling studies. Finally, a practical demonstration of the benefits of aligned porosity on piezoelectric energy harvesting capabilities of ferroelectric ceramics is presented by subjecting poled porous and dense barium titanate materials to mechanical vibrations and charging a capacitor using the harvested electrical energy.

## 2. Single pore model

Finite element analysis has been used to demonstrate the effect of pore shape and orientation on the electric field distribution during the poling process of a ferroelectric ceramic. A single elliptical pore with a constant area fraction ($A_f$ = 0.0785) and relative permittivity, $\varepsilon_r$ = 1, was placed within a high permittivity barium titanate matrix, $\varepsilon_r$ = 1500 (measured experimentally). The angle of the pore was varied from 0° (i.e. parallel to the field) to 90° (perpendicular to the field) and the aspect ratio varied from 1 to 10 (with an aspect ratio of one corresponding to a circular pore). An electric field greater than the coercive field of the barium titanate was applied to top and bottom boundary lines of the model to simulate the poling electric field across the material. The electric field in each element was analysed to determine whether the local field exceeded the coercive field ($E_c$ = 0.5 kV/mm [37]), i.e. the necessary field to switch the orientation of a domain in a ferroelectric material. The applied field used in this model was selected as it provided contrast between poled and unpoled regions; increasing the applied field (analogous to increasing the poling voltage) results in higher poled fractions of material, however, in real materials this is limited by electrical breakdown in the pores that discharges the poling field [38].

An example electric field contour plot for a pore orientated at 45° to the field with aspect ratio = 4, is shown in Fig. 2a, and the resulting distribution of poled material for a range of pore angles is shown in Fig. 2b. It can be seen from Fig. 2a that the electric field concentrates in the low permittivity pore and low electric field regions are present in the $BaTiO_3$ phase in immediate vicinity of the pore at pore-ceramic interfaces perpendicular to the applied field. This results in varying fractions of the $BaTiO_3$ becoming poled as shown in Fig. 2b, in which the blue areas are those in which the local electric field is below the coercive field and is therefore likely to remain unpoled. The resulting fraction of poled material as a function of pore angle with respect to field is shown in Fig. 2c. High aspect ratio pores aligned parallel to the applied field yield the highest fraction of poled material (angle = 0° in Fig. 2b and c), whereas high aspect ratio pores aligned perpendicular to the poling field (angle = 90° in Fig. 2b and c) resulted in the lowest poled fraction. If we were to extend the aspect ratio further we would reach the extreme case of parallel and series structures, i.e. a 3-1 or 2-2 composite in three dimensions (see Fig. 1). These results indicate the potential for achieving high degrees of poling in porous ferroelectric structures by having pores with a high aspect ratio and aligning them parallel to the poling axis, such as those achieved by the freeze casting process; this is discussed in more detail in Section 3.1.

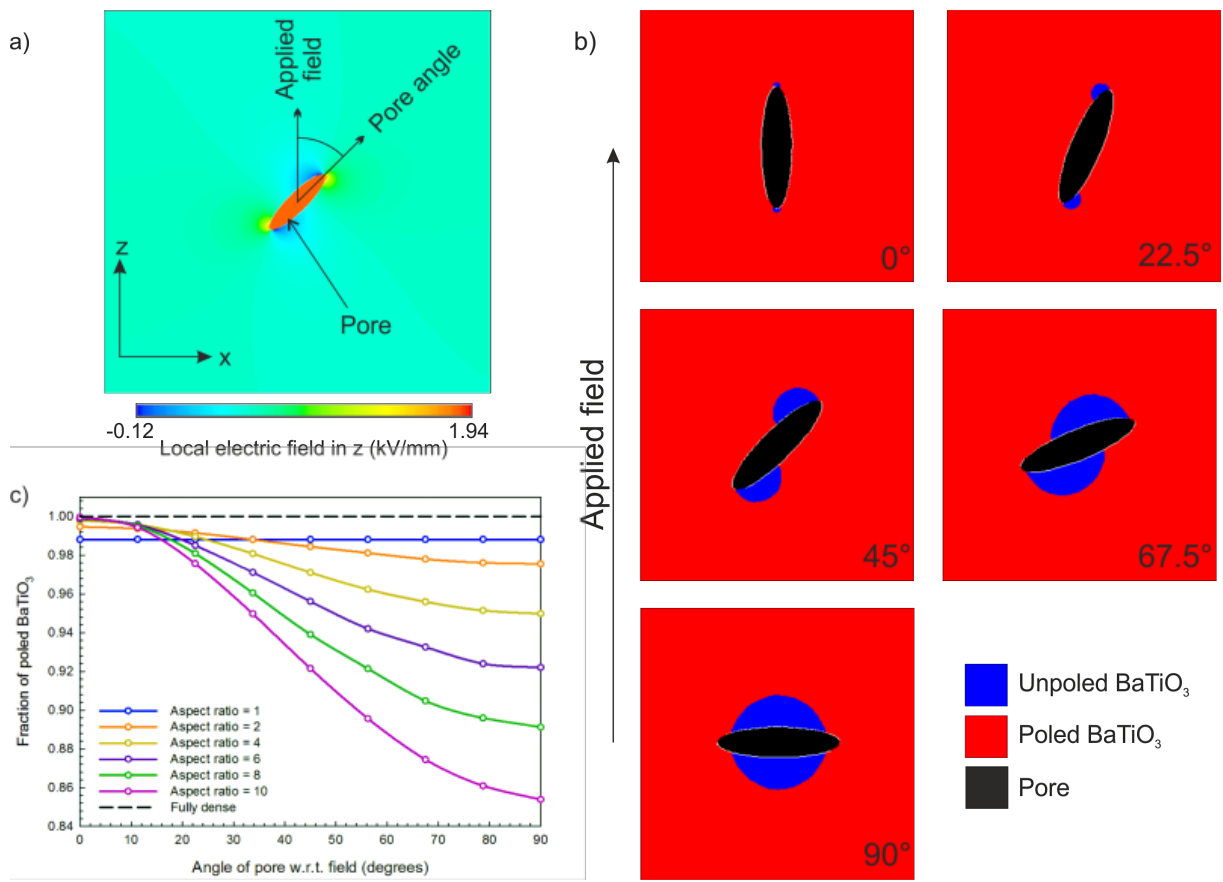

*Figure 2: a) Contour plot of electric field distribution in and around a single pore (aspect ratio, AR = 4, angle = 45°) contained within a barium titanate matrix; b) regions of poled (red) and unpoled (blue) barium titanate with single pore (black) angle varied from 0° to 90° with respect to applied field; and c) effect of aspect ratio and angle of pore with respect to direction of applied field on the fraction of barium titanate poled in z-direction, i.e. the area in which the local field exceeds the coercive field of barium titanate ($E_c$ = 0.5 kV/mm [37]).*

### 3. Porous network model

We now expand upon the single pore model presented in Section 2 to investigate three dimensional aligned structures that better represent the type of microstructure formed via unidirectional freeze casting of water-based solutions. A porous network finite element approach has been implemented that has been used previously to describe the properties of lead zirconate titanate (PZT) with uniformly distributed porosity [9] and barium titanate with porous sandwich layer structures [10]. Firstly, however, it is necessary to understand the freeze casting process and the type of porous structures formed in more detail.

*3.1 The freeze casting process*

Freeze casting is a process whereby a liquid suspension of a solvent and ceramic particles is frozen before the solvent phase is sublimated under reduced pressure and low temperature, i.e. converted from the solid directly to the vapour phase. As the solvent freezes the ceramic is ejected from the ice front forming regions of compacted ceramic powder. When the frozen solvent is sublimated pores are left that have the morphology of the solvent crystals, hence the process often being referred to as 'ice-templating' [39, 40]. The remaining ceramic powder compact is then sintered at high temperature to densify the ceramic walls. A variety of porous structures can be achieved through control of the freezing conditions and the properties of the liquid suspension, however, the majority of work to date in terms of freeze cast porous ceramics has focussed on forming highly aligned structures that are beneficial in terms of their mechanical properties [36]. Similarly, the literature on freeze cast porous ferroelectric ceramics focusses on the fabrication and properties of porous structures aligned to the poling direction that, as previously discussed, are found to have high piezoelectric strain coefficients [7, 22-24]. The excellent alignment is achieved by unidirectional freezing of a suspension, which encourages the crystallising solvent to grow preferentially along the direction of the temperature gradient.

The solvent used in the liquid suspension is a key factor in determining the final porous structure of the freeze cast material. Water is a commonly used solvent and is the focus of the porous structures examined in this work, although tetra-butyl alcohol and camphene have also been extensively researched and the work here also has relevance to the properties of aligned structures formed using these solvents [40]. The crystal structure and the crystal growth kinetics of the solvent determine the pore morphology of the freeze cast material. For example, ice has a hexagonal crystal structure and highly anisotropic growth kinetics so that crystals grow preferentially along their basal plane with limited dendritic growth perpendicular to this plane [41]. During unidirectional freezing the basal plane of the ice crystals are aligned to the freezing direction, in effect growing in a two-dimensional sheet. As the ice grows during water-based freeze casting the ceramic particles are ejected from the solidification front into channels between the sheets of ice, thus forming the characteristic lamellar 2-2 structure [7, 21, 39, 40] (see Fig. 1 for schematic representation of 2-2 structure, Fig. 1d).

The temperature gradient provides the driving force for the directional growth of the crystallising solvent, meaning that high temperature gradients are favourable for producing highly aligned structures [40]. When the gradient and the driving force for directional growth, is too low the solvent crystals are more likely to tilt away from the primary freezing axis, resulting in reduced alignment [39].

Double-sided freezing set ups, whereby suspensions are frozen from both the top and bottom surfaces simultaneously can further improve alignment [21, 42].

The width of the lamellar ice channels can also be controlled by adjusting the freezing rate, with finer pore structures achieved using faster rates [39]. Reducing the pore width of the freeze cast materials has been shown to improve their mechanical strength compared to coarse structures at the same porosity [36]. However, freezing too fast leads to entrapment of particles within the ice, resulting in little or no pore alignment or long range order. Using a freezing rate that is too slow can lead to unstable freezing conditions, causing lateral growth of ice crystals between adjacent lamellae [43]. Some entrapment of particles within the ice is likely to occur during freezing, which may form characteristic bridges between the ceramic lamellae that are commonly observed in freeze cast structures [7].

The main factor controlling the final porosity of the material is the solid loading of the liquid suspension. Freeze casting has been used to create both highly dense ceramics (<1 vol.% porosity), using very low freezing rates and high solid loading (~50 vol.% solid loading), and high porosity materials, such as aerogels (>90 vol.% porosity) [40]. As the solid loading is increased the viscosity of the suspension increases and redistribution of the ceramic particles away from the solidification front becomes more difficult [44]. This may affect the long range order of the porous structure post-sintering, as well as increasing the likelihood of air bubbles becoming trapped during the freezing process, which cause spherical pores in the sintered ceramic.

### 3.2 Modelling procedure

Freeze casting is a complex process and this paper is not intended as a comprehensive review of the research conducted in this field. However, it is important to understand some of the processing challenges that arise during freeze casting in order to model their behaviour in the context of the poling behaviour of a ferroelectric material. As discussed in Section 3.1, control of the freezing process is essential to generate a highly aligned microstructure, and even with relatively good control of cooling rate and particle size some 'defects' (defined in this sense as microstructural features which detract from an ideal 2-2 structure) are likely to be present, such as porosity in the ceramic channels and ceramic bridges that form in the pore channels. These two defect types are commonly observed in water-derived freeze cast structures as shown in Fig. 3a and b.

For the purposes of this modelling study, an idealised structure for a porous water-based freeze cast material can be defined as a lamellar 2-2 structure with no porosity in the ceramic channels and no ceramic bridges in the pore channels. The model enables an examination of the poling behaviour in regions of the material whereby the pore and ceramic channels are aligned parallel to one another; a transverse image of freeze cast barium titanate (i.e. freezing direction is out of plane) is shown in Fig. 3c with the boxed regions indicating where lamellar ceramic/pore channels are parallel to one another.

Two kinds of defects were introduced into the model in varying fractions to understand the effects of these features on the poling behaviour of the freeze cast materials: (i) pores in the ceramic lamellae that form in freeze cast materials due to incomplete compaction or sintering of the ceramic powder on rejection from the solidification front of the solvent (Fig. 3a); and (ii) ceramic elements that are present in the pore channels, i.e. the ceramic bridges that form during freezing casting due to engulfment of particles by the ice front and unstable growth conditions (Fig. 3b). As with the single pore model the finite element analysis was conducted in Ansys APDL.

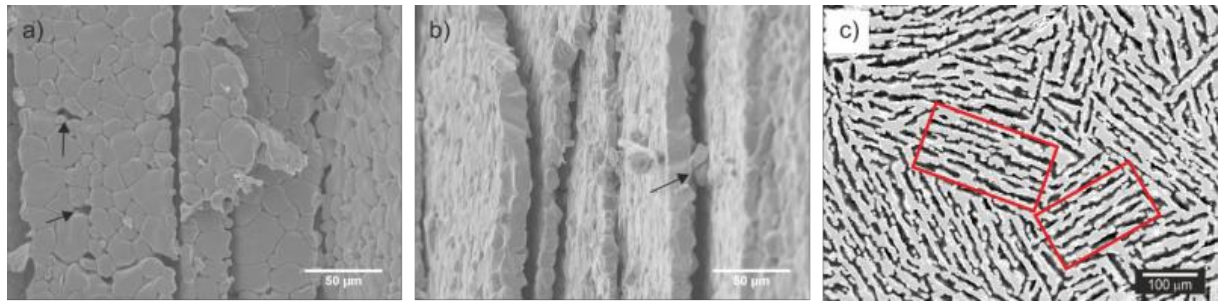

*Figure 3: SEM images of porous barium titanate fabricated via unidirectional freeze casting with 30 vol.% solid loading in suspension (see Section 4.1 for more details) with examples of (a) porosity in the ceramic channels and (b) a bridge between adjacent ceramic lamellae, indicated in both images by arrows; the freezing direction is vertical in both images; and (c) SEM of porous barium titanate taken transverse to freezing direction (pores with dark contrast) with boxes indicating regions of freeze material whereby pore and ceramic lamellae are parallel, such as those modelled in Section 3.*

### 3.3 Definition of the model geometry

A 30 x 30 x 30 mesh of cubic elements was used as the starting point for all structures (Fig. 4a) before designating pore channels (Fig. 4b). Bridges between and porosity within the ceramic channels was then introduced in randomly selected positions within the network (Fig. 4c). The fraction of porosity introduced into the ceramic channels was controlled by the parameter $\alpha_{cc}$ and the fraction of ceramic elements in the pore channels (i.e. the bridges) was controlled by the parameter $\alpha_{pc}$. The parameters are defined such that an ideal 2-2 structure with no porosity in the ceramic channels or ceramic in the pore channel had both $\alpha_{cc}$ and $\alpha_{pc}$ equal to zero, i.e. the structure in Fig. 4b. The variables $\alpha_{cc}$ and $\alpha_{pc}$ were varied from zero to 0.3 and 0.4, respectively. This range was selected as it exceeds the desirable levels of ceramic channel porosity and pore channel bridges in a freeze cast material and was observed to give a good indication of the effect of the two types of 'defect' on the poling behaviour and piezoelectric properties of freeze cast ferroelectric ceramics. Fig. 5 demonstrates the effect of varying $\alpha_{cc}$ (top row) and $\alpha_{pc}$ (bottom row) on porous structure, whereby the images are 2D slices taken parallel to the poling direction.

The pore channel width was maintained constant at one element wide as the model indicated that increasing it did not significantly affect the poling behaviour and would have meant the presence of a ceramic element in the pore channel did not necessarily constitute a bridge between neighbouring lamellae. The spacing between the pore channels (i.e. the ceramic channel width) was varied to provide additional control over the porosity, as in Fig. 5, where the top row of images have a ceramic channel width of three elements and the bottom row of images have a width of two. Once the porous network had been generated the ceramic elements were assigned the properties of unpoled barium titanate (elastic modulus = 120 GPa, Poisson's ratio = 0.3, $\varepsilon_r$ = 1187.5 [46]) and pore elements were assigned the properties of air (elastic modulus = 0, $\varepsilon_r$ = 1). The mesh size was deemed to yield a good balance between computing speed and reliability. As networks were generated randomly, five different structures were evaluated for each combination of $\alpha_{cc}$ and $\alpha_{pc}$ parameters.

The poling procedure was simulated to achieve a distribution of poled and unpoled barium titanate. Electrodes were 'applied' by coupling the voltages of the nodes at the top and bottom surfaces and applying a poling field between them, see Fig. 4c. Elements in which the local electric field exceeded the coercive field, i.e. $E_f > E_c$, were assigned the properties of poled barium titanate [37], as shown in Fig 4d in which red elements are those that have been poled and blue elements are those that remain unpoled. Once the distribution of poled material had been established the effective piezoelectric properties of the network were measured.

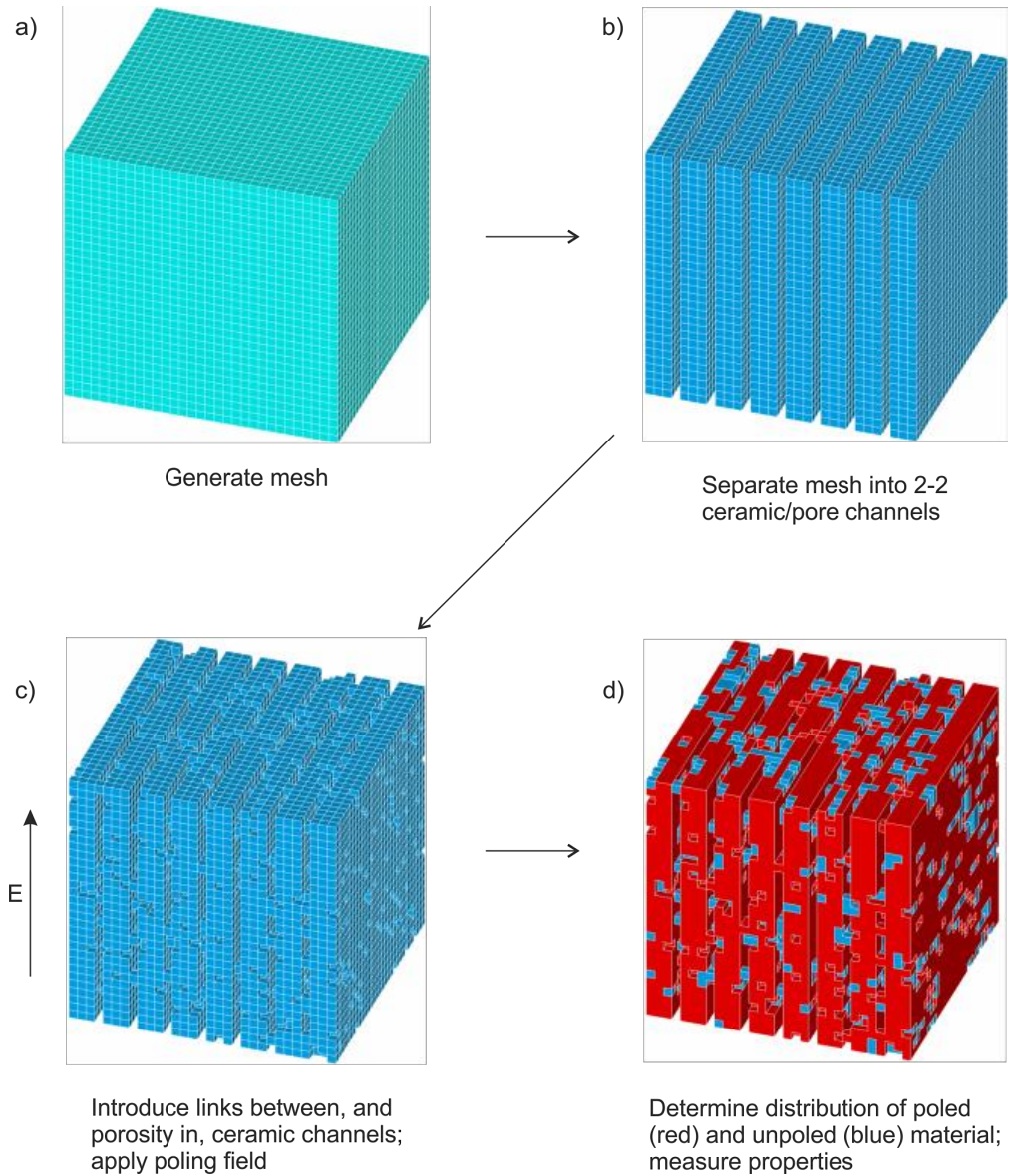

Figure 4: Schematic of process to form porous network geometry in Ansys. a) A cubic mesh with $30^3$ elements was initially generated before b) defining pore channels (i.e. an ideal 2-2 structure), c) introducing porosity in the ceramic channels and ceramic bridges in pore channels and d) applying a poling field to establish the distribution of poled and unpoled BaTiO$_3$ and porosity, and measuring the effective material properties.

*3.4 Porous network model results*

The results for the longitudinal piezoelectric strain coefficient are shown in Fig. 6 for varying the ceramic fraction in the pore channel, $\alpha_{pc}$, with a constant ceramic channel porosity, $\alpha_{cc}$, and Fig. 7 for varying $\alpha_{cc}$, with a constant $\alpha_{pc}$. The data has been fitted using second order polynomial functions to highlight general trends. Firstly, considering Fig. 6, increasing the fraction of barium titanate in the pore channels (increasing $\alpha_{pc}$) resulted in a decrease in the measured $d_{33}$ at a given porosity; the highest $d_{33}$ values were achieved in structures with well-defined pore channels parallel to the poling (z) direction. This was due to regions of low electric field in the immediate vicinity of a ceramic-pore interface perpendicular to the applied field (as can be seen in the single pore model, Fig. 2), so that BaTiO$_3$ ceramic elements in the pore channels tended to be in low field regions and therefore remained unpoled, thus adding to the relative density of the material but not contributing to the measured $d_{33}$. The results shown in Fig. 7 are more intuitive in that increasing porosity within the ferroelectric channels (increasing $\alpha_{cc}$) led to lower poled fractions that resulted in reduced $d_{33}$ coefficients. A linear relationship between $d_{33}$ and fraction of barium titanate poled was observed for all $\alpha_{cc}$ and $\alpha_{pc}$ values, see Fig. 7e, demonstrating a clear link between fraction poled and the resulting piezoelectric response, as reported previously [9, 10].

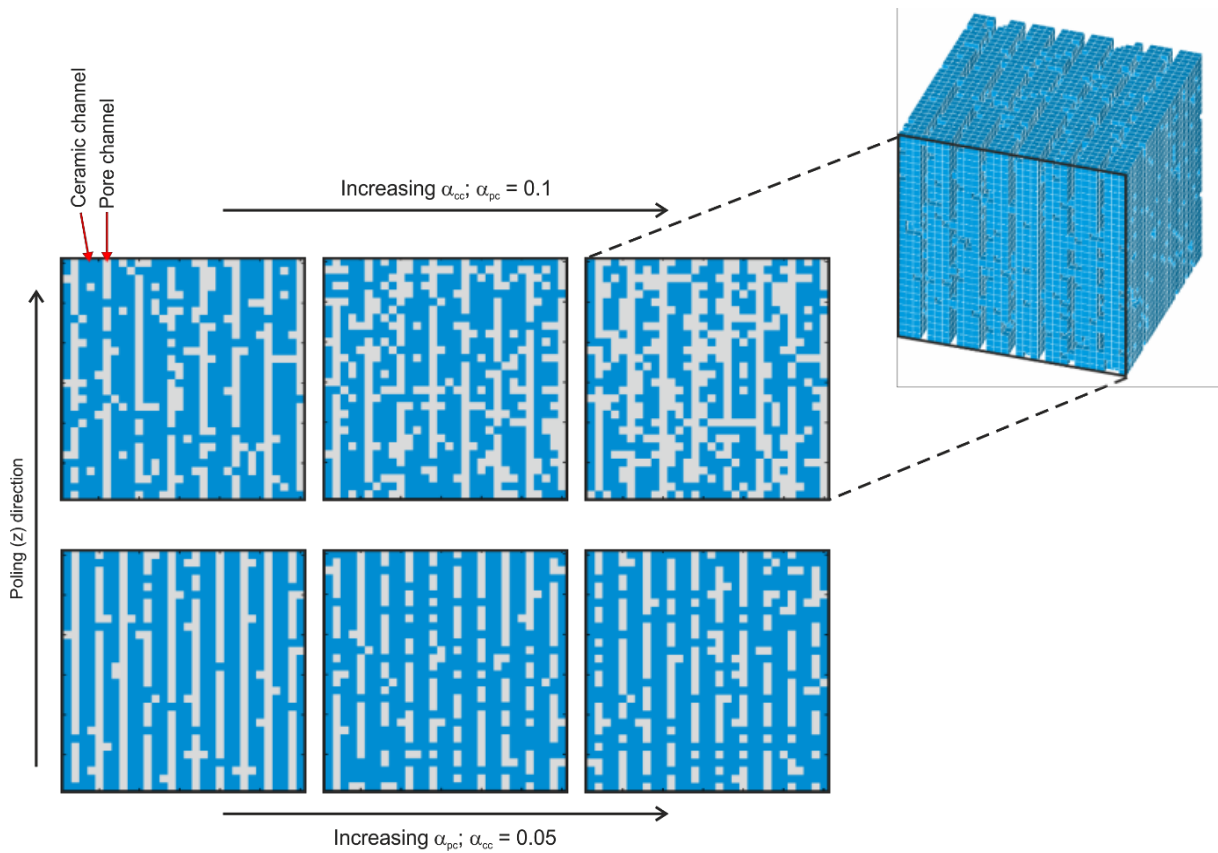

*Figure 5: Cross-section images (see top right for reference) of modelled porous structures showing the effect of the variables, $\alpha_{cc}$, i.e. the fraction of porosity in the ceramic channels, and $\alpha_{pc}$, the fraction of ceramic in the pore channels on the porous structure; pores are light grey in colour and the ceramic phase is blue. The top three images have increasing $\alpha_{cc}$ with $\alpha_{pc}$ = 0.1 and the bottom three images have increasing $\alpha_{pc}$ with $\alpha_{cc}$ = 0.05. The ceramic channel width is wider in the top three images compared to the bottom three images, which allowed additional control over the overall porosity fraction.*

A comparison between model predictions of an ideal 2-2 structure ($\alpha_{cc}$, $\alpha_{pc}$ = 0), freeze cast structures (with varying ceramic channel porosity, $\alpha_{cc}$, and ceramic in pore channel, $\alpha_{pc}$), and uniform porous BaTiO$_3$ is shown in Fig 7. For an ideal 2-2 porous structure, $d_{33}$ remained at 140 pC/N regardless of the porosity fraction, since the highly aligned structure promoted a homogeneous poling field so that all the BaTiO$_3$ elements were poled. At low $\alpha_{cc}$ and $\alpha_{pc}$, i.e. close to an ideal 2-2 structure, the predicted $d_{33}$ was higher than that of barium titanate with uniformly distributed porosity of the same density, as has been shown experimentally by comparing highly aligned freeze cast porous PZT and PZT with uniform porosity manufactured by the BURPS process [7, 15, 16, 22-24]. However, when $\alpha_{pc}$ was increased and the pore channels became less well-defined due to higher fractions of BaTiO$_3$, higher $d_{33}$ coefficients were observed in the uniformly distributed porous structures compared to the freeze cast microstructures, even when there was a low pore fraction within the ceramic BaTiO$_3$ channels (low $\alpha_{cc}$). This indicates that in order to achieve high $d_{33}$ coefficients clear pore channels are more important than highly dense ferroelectric channels. This is possibly caused by a restriction of strain in the poled material due to the presence of unpoled ceramic bridges in the porous channels. For example, when an external field was applied to the model a piezoelectric strain was observed in the highly poled dense channels, but no strain was induced in the stiff unpoled BaTiO$_3$ bridges within pore channels due the applied field. When there was a low fraction of unpoled ceramic in the pore channels (i.e. a low $\alpha_{pc}$) the poled ceramic channels were relatively unrestricted in their piezoelectric expansion due to the applied field so that a high net piezoelectric response was observed throughout the material, leading to a high $d_{33}$. However, when the fraction of ceramic in the pore channel was increased to $\alpha_{pc} > 0.2$ there was sufficient unpoled ceramic bridges in the pore channels to restrict the induced piezoelectric strain in the ceramic channels so that that the net piezoelectric strain of the structure fell below that of the material with uniformly distributed porosity, particularly when $\alpha_{cc}$, i.e. the amount of porosity in the ceramic channels, was also high and there was a lower fraction of poled material in these channels. Having well-defined pore channels and long range order was therefore found to be beneficial in terms of promoting high $d_{33}$ coefficients in the modelled freeze cast porous ferroelectric materials.

Example permittivity data obtained from the porous network model are shown in Fig. 8a for constant pore channel ceramic fraction, $\alpha_{pc}$, with varying pore fraction in ceramic channels, $\alpha_{cc}$ and Fig. 8b for constant $\alpha_{cc}$, varying $\alpha_{pc}$, alongside the permittivity for an ideal 2-2 connected ($\alpha_{cc}$, $\alpha_{pc}$ = 0) and barium titanate with uniformly distributed porosity. The ideal 2-2 structure followed a linear trend of a parallel rule of mixtures model [47] and as $\alpha_{cc}$ and $\alpha_{pc}$ approached zero, i.e. towards an ideal 2-2 structure, the permittivity increased to what is effectively an upper bound. When $\alpha_{pc} > 0.1$ the permittivity of the freeze cast structure became close to that of the uniform porous barium titanate. Fig. 8c is included to show that at the limits of the parameters studied here, i.e. $\alpha_{cc}$ = 0.3, $\alpha_{pc}$ = 0.4, the permittivity of the aligned structure was lower than the uniform porous material. In terms of the energy harvesting figure of merit (Eqn. 1) a low permittivity is beneficial, however, these structures also had relatively low $d_{33}$ coefficients.

Calculated harvesting figures of merit, $d_{33}^2/\varepsilon_{33}^\sigma$, for the modelled data are shown in Fig. 9. Introducing porosity was generally found to increase the figure of merit compared to the dense value of ~1.40 pm$^2$/N, with the ideal 2-2 structure providing an upper limit for the increases that could be achieved by forming highly aligned freeze cast barium titanate. For example, the uniformly distributed porous material was predicted to have an energy harvesting figure of merit of ~2 pm$^2$/N at 50 vol.% porosity, whereas for the ideal 2-2 structure the predicted figure of merit was ~3 pm$^2$/N at the same porosity, double that of the dense material. Having a low fraction of porosity in the ceramic channels ($\alpha_{cc}$) and a low fraction of ceramic in the pore channels ($\alpha_{pc}$) yielded high figures of merit compared to both dense and porous material due to the high piezoelectric strain coefficient, $d_{33}$, that resulted from high fractions of the ferroelectric phase becoming poled, see Fig. 6 and Fig. 7.

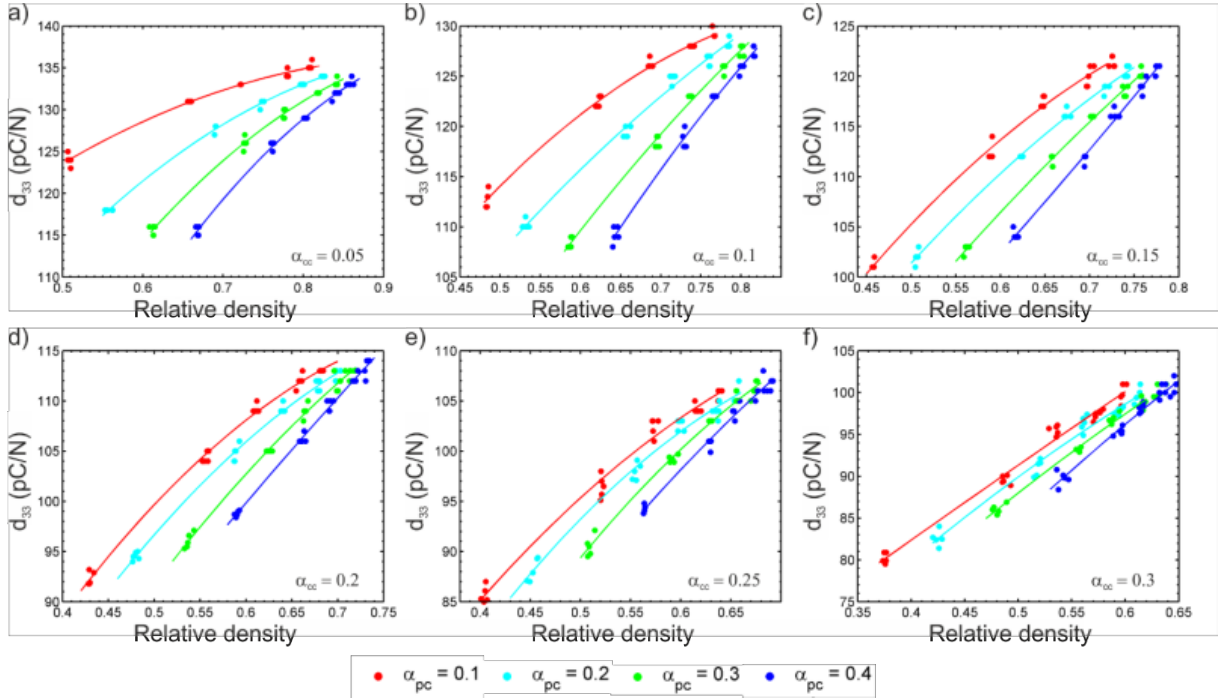

*Figure 6: Effect of fraction of barium titanate in pore channel, $α_{pc}$, on the longitudinal piezoelectric strain coefficient, $d_{33}$, for increasing fractions of porosity in ceramic channel, $α_{cc}$, in a) - e). For a given $α_{cc}$ and $α_{pc}$, the relative density of the modelled material was controlled by adjusting the width of the ceramic channels.*

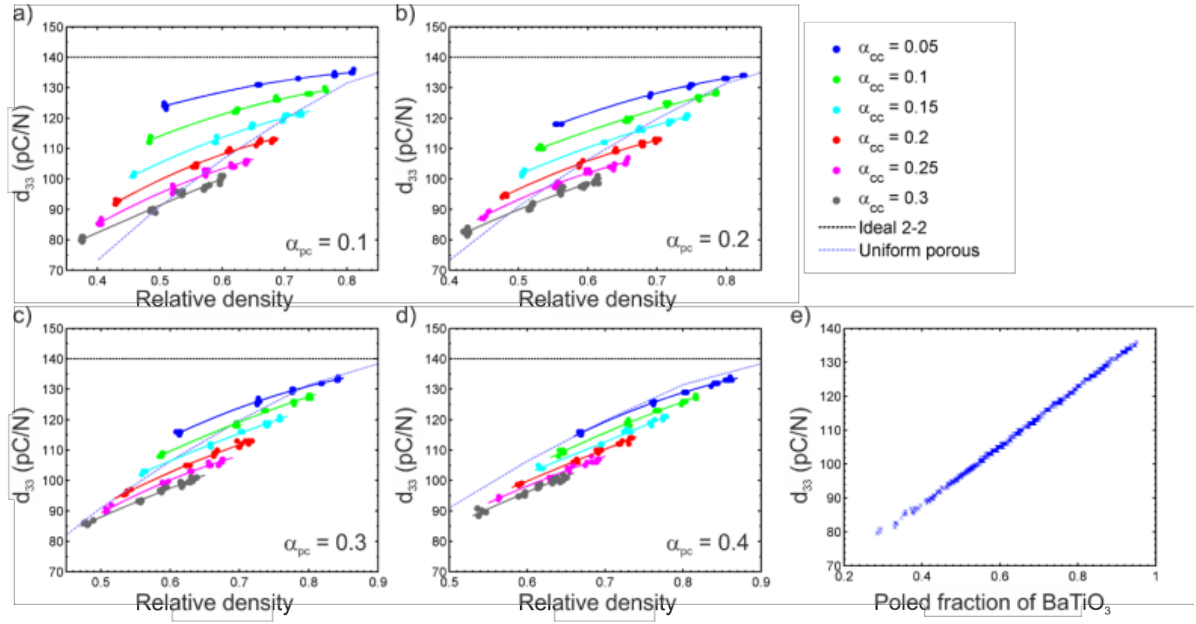

*Figure 7: Effect of pore fraction in ceramic channel, $α_{cc}$, on longitudinal piezoelectric strain coefficient, $d_{33}$, for increasing fractions of ceramic in pore channels, $α_{pc}$, in a) - d) with comparison to an ideal 2-2 structure ($α_{cc}$; $α_{pc}$ = 0) and barium titanate with uniformly distributed porosity, see black and violet dashed lines, respectively; and e) shows the relationship between the fraction of barium titanate poled and the resulting piezoelectric strain coefficient, $d_{33}$.*

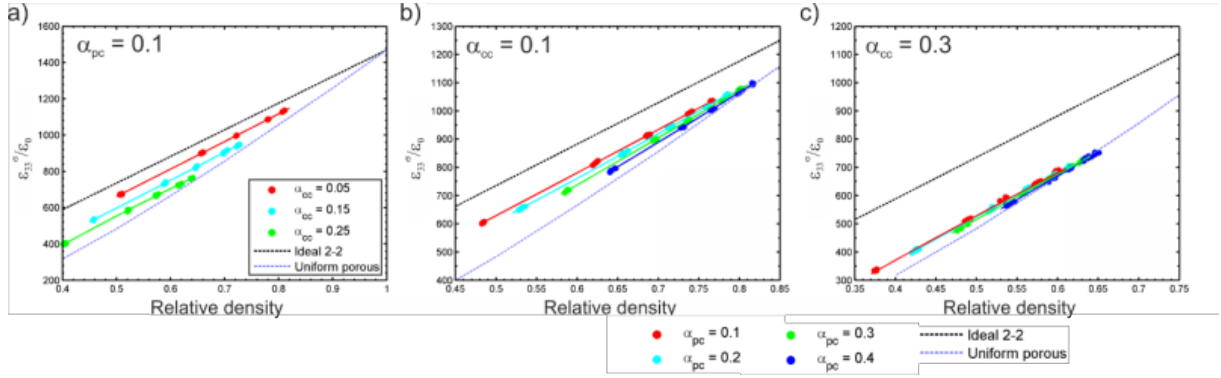

Figure 8: Variation of relative permittivity $\varepsilon_{33}^{\sigma}/\varepsilon_0$ with relative density for a) varying ceramic channel porosity $\alpha_{cc}$, $\alpha_{pc}$ = 0.1, b) varying ceramic fraction in pore channel $\alpha_{pc}$, $\alpha_{cc}$ = 0.1, and c) varying ceramic fraction in pore channel $\alpha_{pc}$, $\alpha_{cc}$ = 0.3. The relative permittivity for an ideal 2-2 structure ($\alpha_{cc}$; $\alpha_{pc}$ = 0) and uniformly distributed porous barium titanate as a function of relative density are shown for comparison (black and violet dashed lines, respectively) in all figures.

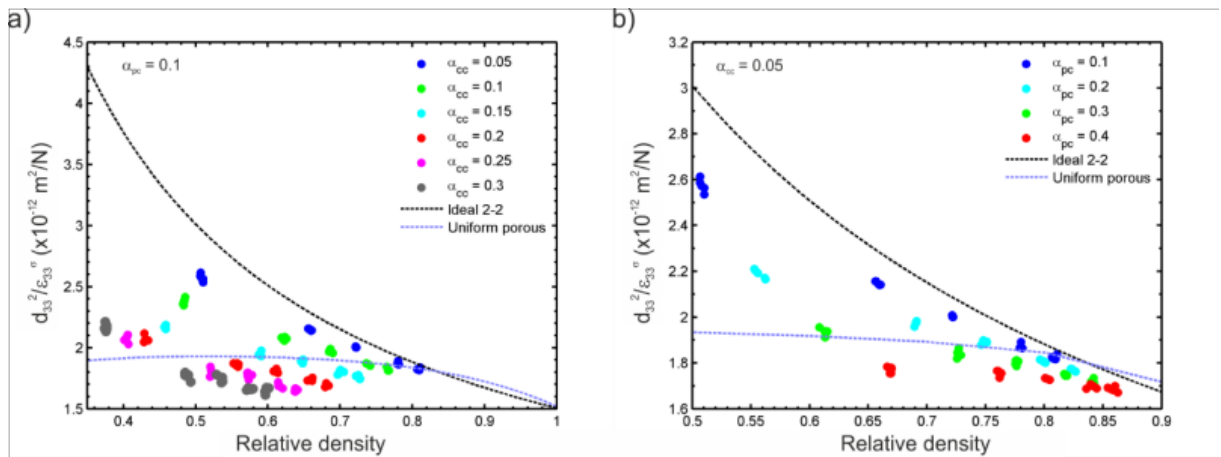

Figure 9: Modelled data showing effect of a) variation of ceramic channel porosity, $\alpha_{cc}$, at constant pore channel ceramic fraction ($\alpha_{pc}$ = 0.1) and b) variation of pore channel ceramic fraction, $\alpha_{pc}$, at constant ceramic channel porosity ($\alpha_{cc}$ = 0.05) on the longitudinal energy harvesting figure of merit of porous barium titanate, $d_{33}^2/\varepsilon_{33}^{\sigma}$. The figure of merit for an ideal 2-2 structure ($\alpha_{cc}$; $\alpha_{pc}$ = 0) and uniformly distributed porous barium titanate as a function of relative density are shown for comparison (black and violet dashed lines, respectively) in both figures.

4. **Experimental**

To provide experimental validation of the finite element model observations, freeze cast barium titanate samples were fabricated with a range of porosities (37 to 56 vol.% porosity). This section details the experimental procedure, results and discussion, and finally a demonstration of the energy harvesting capabilities of the porous ferroelectric materials compared to the dense material, in which samples were mechanically excited and the derived electrical energy used to charge a reference capacitor.

*4.1 Experimental method*

Barium titanate suspensions with varying solid loading contents (20, 25, 30, 32.5 vol.%) were prepared with commercial BaTiO$_3$ powder (particle size, $d_{50}$ = 2.1 µm, Ferro, UK), 1 wt.% organic binder

(polyethylene glycol, Sigma, UK) and dispersant (polyacrylic acid, Sigma, UK), and deionized water. A single-side freeze casting set up was used whereby the suspension was poured into a polydimethylsiloxane (PDMS) mould and placed onto a heat sink cooled by liquid nitrogen to -90°C. The mould used was open ended and aluminium adhesive tape was used to contain the suspension. Care was taken not to introduce air bubbles when filling the moulds. The ice was sublimated from the frozen bodies by freeze drying (Mini Lyotrap, LTE Scientific, UK) for 24 h. These were then sintered at 1300°C for 2 h with a 2 h dwell at 400°C to remove the organic binder; heating and cooling rates during the sintering process were ±60°C/h. Individual samples were cut to ~2 mm thick from the sintered bodies at least 6 mm from the freezing surface. Samples were ground flat, cleaned and silver electrodes applied (RS Components, Product No 186-3600, UK). The ceramic powder and sintering profile were the same as those used in previous investigations [10, 20] to enable comparison between the properties of barium titanate with different porous structures.

The relative density and porosity of the sectioned pellets was measured via the Archimedean method [48]. The samples were corona poled in air at 115°C with 14 kV applied from a 35 mm point source, with the field maintained whilst the rig cooled to 40°C. Piezoelectric strain coefficients, $d_{33}$ and $d_{31}$ were measured 24 h after poling via the Berlincourt method (Take Control Piezometer PM25, UK). Impedance spectroscopy (Solartron 1260 and 1296 Dielectric Interface, UK) was used to measure the permittivity. Microstructural analysis was undertaken using scanning electron microscopy (SEM, JEOL JSM-6480LV).

### 4.2 Microstructural analysis

SEM images of barium titanate freeze cast from suspensions with 20 vol.% initial solid loading (relative density, $\rho_{rel}$ = 0.45) are shown in Fig. 10a and b and 30 vol.% solid loading samples ($\rho_{rel}$ = 0.55) are shown in Fig. 10c and d. Comparing Fig. 10a and c it can be seen that the barium titanate freeze cast from the 30 vol.% loading suspension had more defined pore and ceramic channels than the samples cast from the 20 vol.% suspension. One possible reason for this is reducing the solid loading content resulted in more shrinkage during sintering (30% radial shrinkage for 20 vol.% suspension, 24% radial shrinkage for 30 vol.% suspension) that may have resulted in coalescence of adjacent ceramic lamellae. In addition, adjusting the solid loading of the suspension can alter the freezing temperature and viscosity [40] and therefore it may be that a more aligned freeze cast structure could have been achieved if the freezing conditions were altered depending on the suspension properties. The grain size, which can be seen in Fig. 10c and d, did not change significantly with varying solid loading content or porosity level, with grain sizes generally between 10 and 30 µm, which was similar to previous studies into the effect of porous structure on the energy harvesting properties of barium titanate [10, 20].

Transverse images, i.e. perpendicular to freezing direction, in Fig. 11 and polished images of sample cross-sections (i.e. the direction shown in Fig. 10) were used to calculate the average ceramic and pore channel width of the freeze cast barium titanate. Transverse pore structures of three solid loadings (20, 30 and 32.5 vol.%) with corresponding relative densities of $\rho_{rel}$ = 0.45, 0.55 and 0.62 are shown in Fig. 11a-c, respectively. Increasing the solid loading content led to an increase in ceramic channel average width, which was found to be ~9 µm for barium titanate cast from 20 vol.% suspension compared to ~15 µm for 30 and 32.5 vol.% initial solid loading. On comparing the average ceramic wall thickness measured from transverse images in Fig. 11 to the SEM images in Fig. 10, it appears the ceramic channels were generally 1-2 grains thick with a degree of anisotropy of the grains as the ceramic channel width was smaller than the typical grain size observed when imaging side-on to the ceramic walls, as in Fig. 10b and d; this indicates that freeze casting may also lead to some degree of texturing within the ceramic phase that may be of interest to study further. Pore channel width peaked for the 30 vol.% barium titanate at ~14 µm, with both the 20 and 32.5 vol.% material having a pore channel width of <10 µm although it is thought this is due to differing mechanisms. Firstly, as previously discussed, a low solid loading sample experiences higher shrinkage during sintering, which may have led to a reduction in the pore channel width as neighbouring ceramic lamellae shrink closer together during sintering. The samples cast from suspensions with a high solid loading content had less solvent and more ceramic, which is likely to lead to relatively thicker ceramic channels compared to the pore channels.

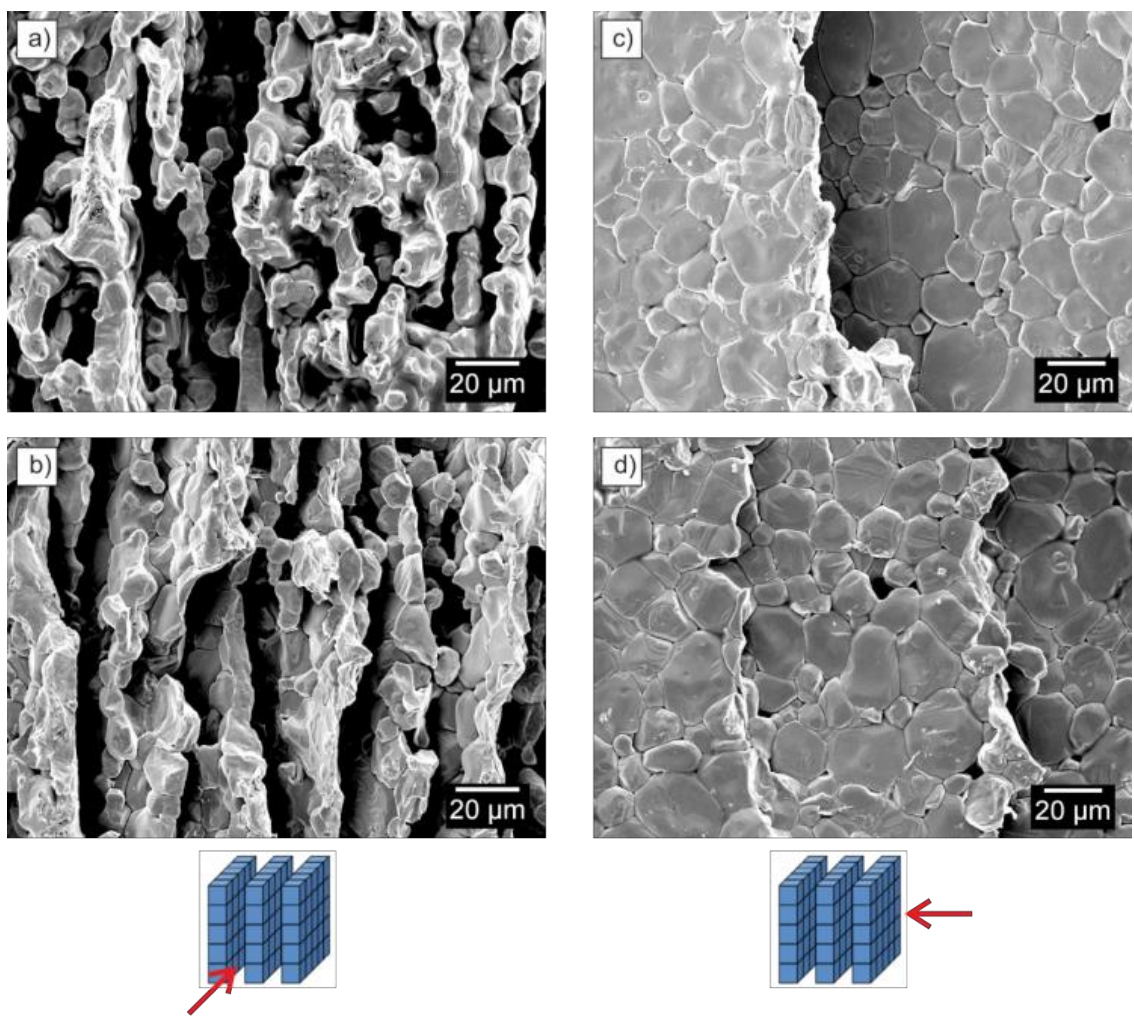

Figure 10: SEM images of freeze cast barium titanate with a) 55 vol.% and b) 45 vol.% porosity ($\rho_{rel}$ = 0.45 and 0.55, respectively; initial solid loading 20 and 30 vol.%, respectively); images c) and d) were taken from samples with 55 and 45 vol.% porosity, respectively, showing grain size did not change with porosity or solid loading content of frozen suspension. The freezing direction was vertical in all images; schematics of the image direction with respect to the freeze cast structure are shown below the SEM images.

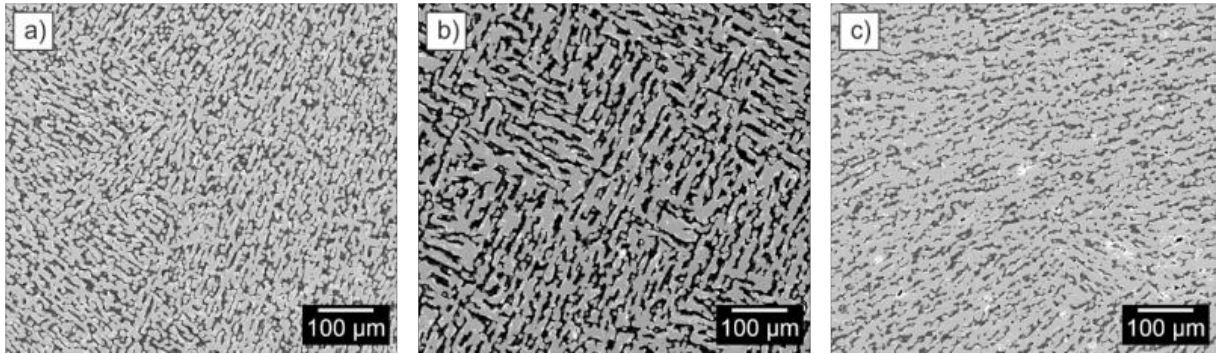

Figure 11: Transverse SEM images (perpendicular to freezing direction) of barium titanate freeze cast with varying solid loadings and different resulting porosities, a) 20 vol.%, $\rho_{rel}$ = 0.45, b) 30 vol.%, $\rho_{rel}$ = 0.55 and c) 32.5 vol.%, $\rho_{rel}$ = 0.62.

*4.3 Results and discussion*

The measured $d_{33}$ data for freeze cast barium titanate are plotted in Fig. 12a as a function of relative density alongside data for uniformly distributed porous (3-3) barium titanate manufactured via the BURPS process [20]. The $d_{33}$ coefficients of the freeze cast materials were consistently higher than materials with uniformly distributed porosity, with a maximum of 134.5 pC/N measured in barium titanate with a porosity of 45 vol.% (relative density, $\rho_{rel}$ = 0.55), which was 93% that of the dense BaTiO$_3$, where the maximum measured $d_{33}$ was 144.5 pC/N, and can be seen to be higher than the $d_{33}$ of the uniformly distributed porous samples across the whole range of relative densities. This was a similar trend to that observed in the finite element models, see Fig. 6 and 7, in which the highly aligned structures had superior piezoelectric strain coefficients to the uniformly distributed porous samples. The variability in the data indicates there was some variation occurring during the processing that results in some samples having better pore channel and ceramic phase alignment than others, as observed in the results of the modelling study in Section 3, which can significantly affect the piezoelectric properties. The slight decline in $d_{33}$ observed as relative density increased beyond ~0.55 may be a result of increasing fractions of ceramic in the pore channels, which was shown to be detrimental to the longitudinal piezoelectric properties in the porous network model, see Fig. 6 and 7.

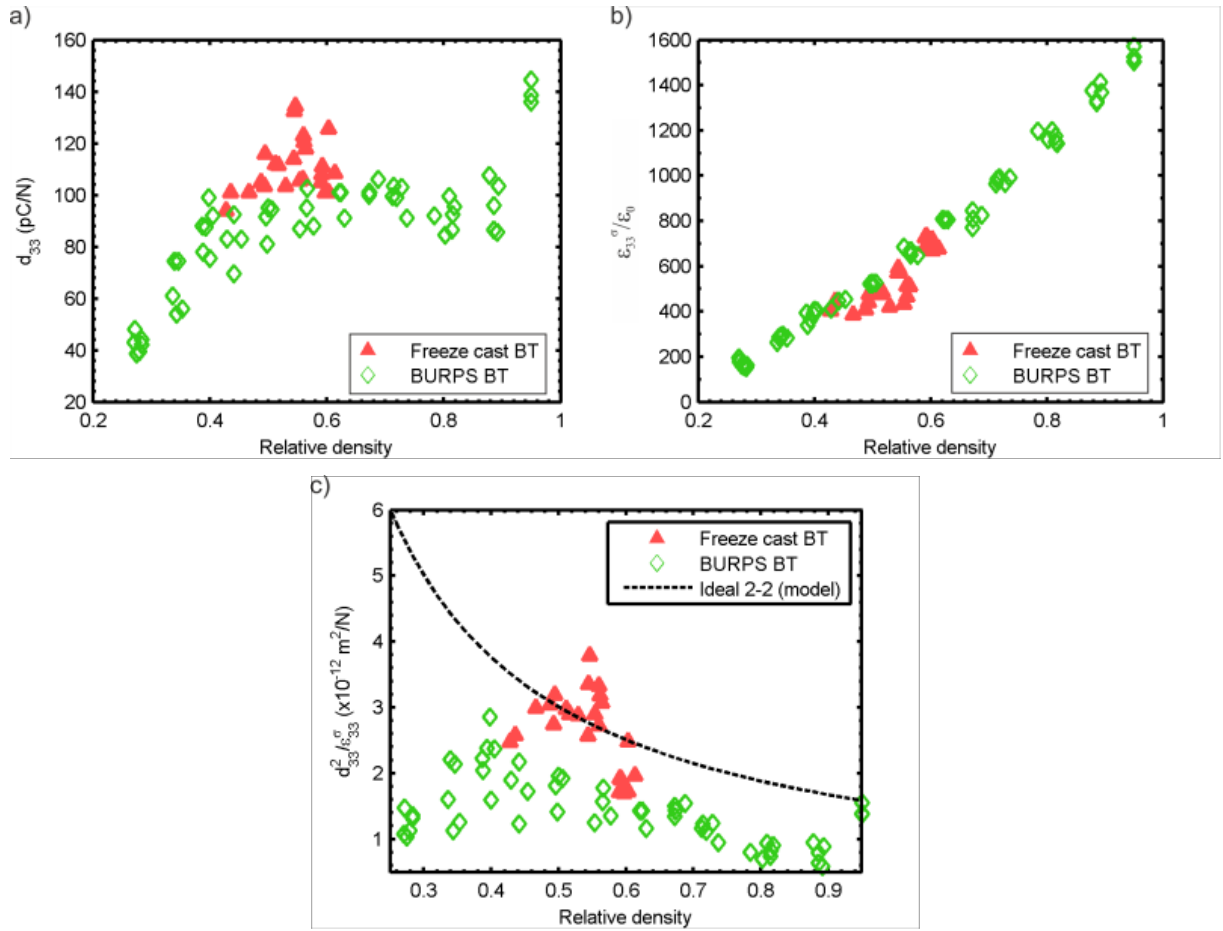

*Figure 12: Comparison of freeze cast (red) and uniform porous (green, manufactured using burned out polymer spheres (BURPS) process, from [19]) for a) longitudinal piezoelectric strain coefficient, $d_{33}$, b) relative permittivity, $\varepsilon_{33}^{\sigma}/\varepsilon_0$, and c) energy harvesting figure of merit ($d_{33}^2/\varepsilon_{33}^{\sigma}$) as a function of relative density.*

Relative permittivity at 1 kHz is shown in Fig. 12b as a function of relative density. It can be seen that the aligned freeze cast barium titanate had a slightly lower permittivity than the uniform porous barium titanate, which was not predicted by either finite element model presented in this paper or indeed other models discussed elsewhere [47, 49]. However, these models assume there is no physical change in the material properties with the introduction of porosity or processing conditions as the model input properties are those measured from the fully dense ceramic. As discussed in Section 4.2, the grains in the freeze cast material appeared to be elongated in the freezing direction, whereas the barium titanate formed via the BURPS process that the freeze cast permittivity data are compared to in Fig. 12b had equi-axied grains [20] and it is known that the microstructure can alter significantly the properties of barium titanate. For example, the dielectric properties of barium titanate are influenced by grain size, with permittivity decreasing as grain size increases, which is thought to be a result of changing domain wall mobility with grain size [50-52]. As an example, single crystal barium titanate has highly anisotropic dielectric properties [53] with lower permittivity in the poling direction compared [37] and texturing the material can also affect the measured permittivity [54].

While any grain size or texturing effects are also likely to alter the piezoelectric properties of barium titanate [52, 54, 55], such as $d_{33}$, the model and experimental data suggests that the piezoelectric coefficient is more dependent on the poling behaviour due to the porous structure, which may explain why the experimental $d_{33}$ data in Fig. 12a is closer to the behaviour as predicted by the model. Further work is required to fully understand the experimental permittivity data, however, it is of interest to note that reduced permittivity (compared to the finite element model presented here and rule of mixtures model for a parallel-connected structure [47]) are beneficial in terms of the energy harvesting figure of merit, see Eqn. 1.

The high $d_{33}$ coefficients measured in the freeze barium titanate samples led to significantly increased energy harvesting figures of merit compared to uniformly distributed porous barium titanate, see Fig. 12c, and when combined with the reduction in permittivity yielded a two-fold increase in harvesting figure of merit compared to the dense material. A maximum of 3.79 pm$^2$/N was found in the freeze cast barium titanate at a relative density of 0.55 and many samples were found to have a figure of merit of >3.0 pm$^2$/N, more than twice that of dense barium titanate. This was higher than was predicted by the finite element model due to low permittivity in the experimental results; the model data for an ideal 2-2 structure are shown in Fig. 12c (black dashed line) for comparison with the experimental data.

### 4.4 Piezoelectric energy harvesting demonstrator

A piezoelectric energy harvesting system, shown in Fig. 13, was used to demonstrate the improved harvesting of the porous freeze cast barium titanate compared to the dense material with three samples selected for comparison: a dense barium titanate pellet with a figure of merit of 1.39 pm$^2$/N, and two freeze cast samples, with a relative density of 0.55 and 0.60 and harvesting figure of merit of 3.24 pm$^2$/N and 2.19 pm$^2$/N, respectively; see Table 1 for more details. The samples all had a thickness of ~1.8 mm and a diameter of ~10 mm to keep the volume similar. The piezoelectric samples were fixed to a Perspex beam attached to a shaker (LDS V201, Bruel & Kjær, DK) with conductive silver epoxy (CircuitWorks Conductive Epoxy, Chemtronics, USA). The perpendicular distance from the central axis of the shaker to the location of the sample was minimised so as not to induce any bending effects. An end mass weighing six grams was attached to the sample with silver epoxy, see Fig. 13a. The samples were excited off-resonance at 97 Hz on the shaker driven at 205 mV by a signal generator passed through an amplifier with the gain set to 20 dB (x10 amplification); calibration using an accelerometer found these conditions to give a maximum acceleration force of approximately 11 $g_0$. The open circuit voltage and short circuit current were measured using a B2987A Electrometer (Keysight, UK). The experimental set up is shown in Fig. 13b. The voltage generated by a sample was then measured again after being passed through a full rectifying bridge circuit with four 1N4148 diodes (NXP Semiconductors, NL), which enabled calculation of the voltage drop across the rectifier. Finally, a 1 µF capacitor (63 V, Philips 030K0, USA) ($C_L$ in Fig. 13c) was charged across a load resistance of 10 MΩ measured with a differential probe of 20 MΩ input impedance, resulting in an equivalent of 6.67 MΩ load resistance ($R_L$ in Fig. 13c). To reduce noise, the tests were conducted in a Faraday cage with shielding and grounding connect to electrometer's guard and chassis ground, respectively. Open circuit voltage and short circuit current experiments were repeated six times for each sample and the data averaged; capacitor charging experiments were repeated three times per sample. The voltage drop across the rectifying bridge was 0.2 V. Capacitances of the samples were measured with the HP 4263B LCR Metter with test signal level of 1V at 100 Hz.

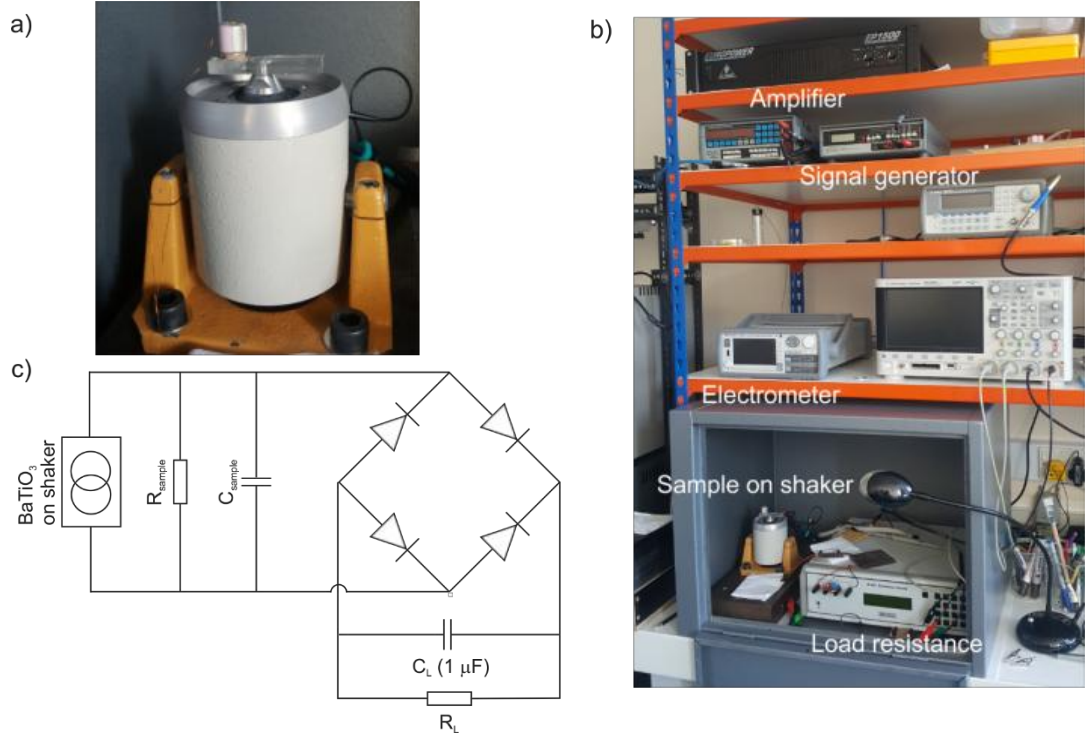

*Figure 13: Experimental setup for piezoelectric energy harvesting experiments; a) sample was attached to a Perspex beam fixed to a shaker with silver epoxy and a six gram end mass attached to the other side, b) shows the full set up with relevant measuring apparatus labelled and c) is the circuit diagram for the rectifying circuit and measurement of voltage across capacitor that was charged using electrical energy converted from input mechanical energy by the barium titanate samples.*

The open circuit voltage, $V_{oc}$, and short circuit current, $I_{sc}$, are shown in Fig. 14a and b, respectively; data sets have been smoothed and averaged. The open circuit voltage increased with increasing porosity, with a maximum peak-to-peak open circuit voltage, $V_{oc}$ = 1.15 V for the barium titanate with $\rho_{rel}$ = 0.55 compared to $V_{oc}$ = 0.62 V for the dense barium titanate. The theoretical open circuit voltage generated by a piezoelectric due to an applied stress, σ, can be calculated from the following equation [56]:

$$V_{oc} = \frac{d_{33}}{\varepsilon_{33}^{\sigma}}.t.\sigma \qquad (3)$$

where $t$ is the sample thickness. The piezoelectric strain coefficient, $d_{33}$, and sample thickness was similar in the three tested samples and so under the same loading conditions $V_{oc}$ was expected to be approximately inversely proportional to the permittivity of the sample, as was observed experimentally. The short circuit current, $I_{sc}$, decreased with increasing porosity and decreasing permittivity, see Fig. 14a and Table 1.

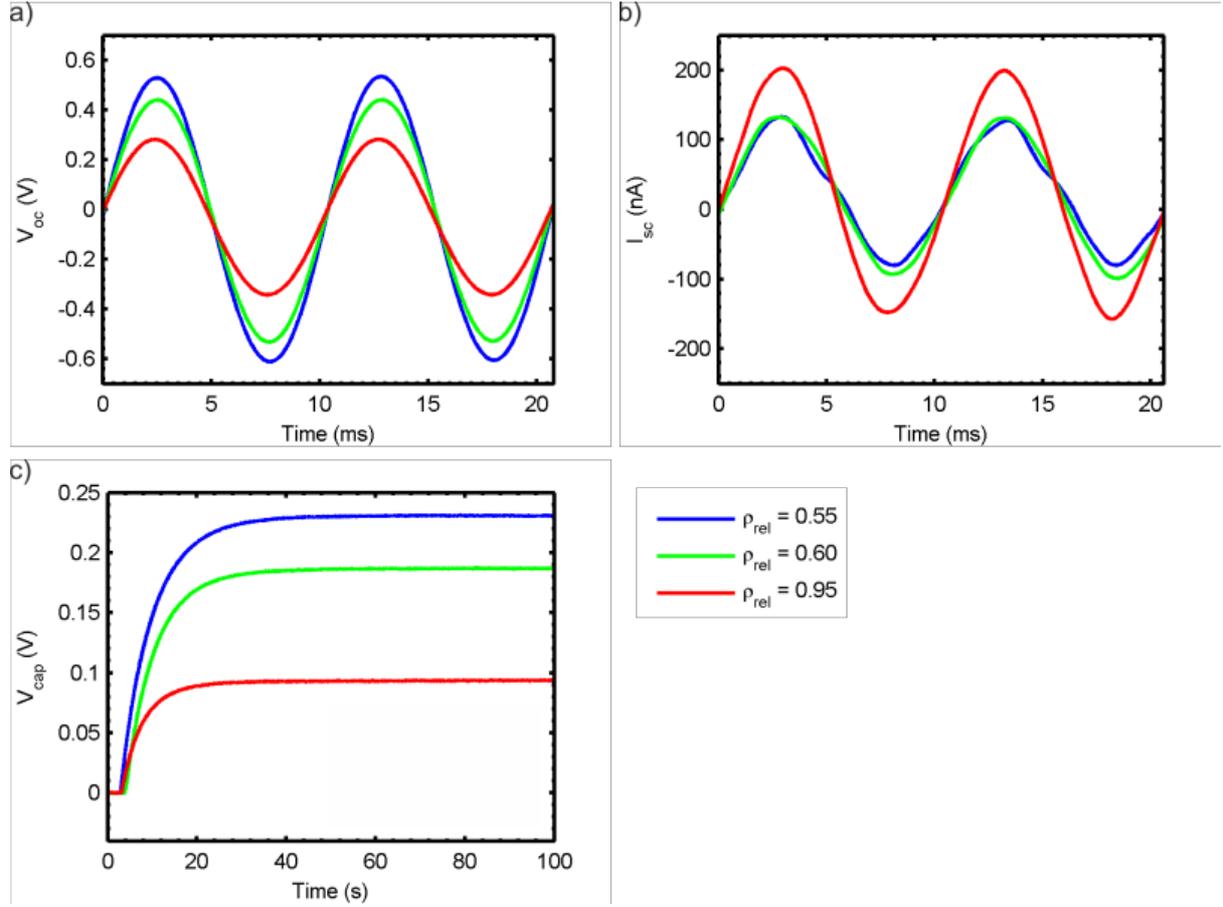

Figure 14: Results from shaker experiment showing comparison between two porous freeze cast barium titanate samples (blue and green lines) and dense barium titanate (red) for a) open circuit voltage, b) short circuit current and c) capacitor charging curve after rectification. Sample details are shown in Table 1.

Table 1: Details of samples used in piezoelectric energy harvesting experiments.

| $\rho_{rel}$ | $d_{33}$ (pC/N) | $\varepsilon_r$ | $d_{33}^2/\varepsilon_{33}^\sigma$ (pm²/N) | $V_{oc}$ (V) (peak-to-peak) | $I_{sc}$ (nA) (peak-to-peak) | $V_{cap}$ (mV) | C (pF) |
|---|---|---|---|---|---|---|---|
| 0.95 | 136 | 1504 | 1.39 | 0.63 | 360 | 96 | 771 |
| 0.60 | 118 | 717 | 2.19 | 0.97 | 231 | 196 | 244 |
| 0.55 | 129 | 580 | 3.24 | 1.15 | 213 | 234 | 184 |

The dense sample had the highest peak-to-peak short circuit current, $I_{sc}$ = 360 nA, compared to the samples with $\rho_{rel}$ of 0.60 and 0.55, which had peak-to-peak short circuit currents of 231 and 213 nA, respectively. The reason for the fall in current with increasing porosity is thought to be an effect of the varying capacitances of the samples. The current produced by the sample due to mechanical excitation, $I$, is equal to the change in charge, $dQ$, with time, $dt$:

$$I = \frac{dQ}{dt} \quad (4)$$

and the change in charge due to a change in voltage, $dV$, can be written as:

$$dQ = C.dV \quad (5)$$

if we assume constant capacitance, C, of the piezoelectric during the experiment. If we differentiate with respect to time and rearrange we get the relationship:

$$I = \frac{dQ}{dt} = C \cdot \frac{dV}{dt}. \tag{6}$$

From Eqn. 6 it can be seen that for a given change in voltage with time, *dV/dt*, the dense sample was expected to produce the most current as it had the highest capacitance, four times greater than the capacitance of the highest porosity sample, see Table 1. However, as the porous samples had a lower permittivity they produced higher voltages per unit stress (see Eqn. 3 and Fig. 14a) therefore the peak current of the dense barium titanate due to an applied stress was not four times greater than the barium titanate with $\rho_{rel}$ = 0.55, as predicted by their respective capacitances, but was only ∼1.7 times greater due to a combination of higher capacitance yielding increased current outputs whilst simultaneously causing a reduction in the voltage generated due to an applied stress.

Fig. 14c shows that the porous barium titanate more effectively charged the capacitor than the dense material with a maximum voltage of $V_{cap}$ = 234 mV achieved for the sample with $\rho_{rel}$ = 0.55, compared to $V_{cap}$ = 96 mV for the dense sample (average $V_{cap}$ from three measurements). The time to achieve the peak voltage in the barium titanate with $\rho_{rel}$ = 0.55 was 40 seconds. The results from these experiments indicate the rate and magnitude of harvested energy increased with increasing porosity, as was predicted by the piezoelectric energy harvesting figures of merit of the three samples tested, demonstrating proof-of-concept for the use of porous ferroelectric materials in piezoelectric energy harvesters. The reduced capacitance of the porous samples compared to the dense materials may also enable easier impedance matching if this experiment was to be developed further into a prototype harvesting device.

## 5. Conclusions

This paper has demonstrated the potential for the controlled introduction of porosity into ferroelectric ceramics as a method to improve their longitudinal energy harvesting capabilities. Two new finite element models were presented that investigated the effect of pore structure on the poling behaviour and piezoelectric properties of porous ferroelectric materials in varying levels of detail, followed by an experimental study whereby highly aligned porous barium titanate was fabricated via the freeze casting method and found to have improved energy harvesting properties compared to dense barium titanate.

Firstly, a single pore model showed that high aspect ratio pores aligned to the poling direction aided the poling of the porous ferroelectric material through promotion of homogeneous poling fields that yield high fractions of poled ferroelectric phase, compared to pores with lower aspect ratios not aligned to the poling axis. This model provides an explanation for high piezoelectric strain coefficients that have been reported in freeze cast ferroelectric ceramics, highlighting the importance of the link between pore morphology and structure and the resulting piezoelectric properties of porous ferroelectric materials.

A second finite element model was then used to more accurately model 2-2 connected porous structures commonly achieved through water-based freeze casting. A porous microstructural network approach was used to investigate the effect of porosity in the ceramic channels and ceramic bridges in the pore channels. Increasing the fraction of porosity in the ceramic channels reduced the fraction of material poled and therefore the effective longitudinal piezoelectric strain coefficient, $d_{33}$ of the material. However, of greater interest was the detrimental effect of the presence of ceramic bridges in the pore channels, which tended to remain unpoled due to their location within the porous structure, whilst acting to restrict the induced strain in the highly poled ceramic channels, leading to lower $d_{33}$ coefficients than similar porosity structures with clear pore channels.

Freeze cast porous barium titanate samples were fabricated and characterised in terms of their piezoelectric and dielectric properties and compared to dense barium titanate and porous barium titanate with spherical, uniformly distributed pores. The freeze cast material had similar $d_{33}$ coefficients as the dense material up to ~50 vol.% porosity, significantly higher than the barium titanate with uniformly distributed porosity. The permittivity of the material was found to decrease with increasing porosity in all cases, leading to excellent piezoelectric energy harvesting figures of merit, with a maximum of 3.79 pm$^2$/N achieved at 45 vol.% porosity (relative density, $\rho_{rel}$ = 0.55), compared to ~1.40 pm$^2$/N for dense barium titanate.

Finally, the benefits of introducing porosity into ferroelectric materials in terms of the piezoelectric energy harvesting performance were demonstrated. Dense and freeze cast porous barium titanate samples were mechanically excited on a shaker and the electrical response used to charge a 1 μF capacitor. The porous samples were found to charge the capacitor at a faster rate than the dense material, as predicted by the energy harvesting figures of merit. The maximum measured voltage across the charged capacitor was found to be 234 mV for freeze cast barium titanate with 45 vol.% porosity, compared with 96 mV for dense barium titanate; corresponding to an improvement of 140%. This work highlights the significant benefits in harvesting by forming ferroelectric microstructures containing highly aligned pores using freeze casting.

**Acknowledgements**

J. I. Roscow would like to thank EPSRC for providing financial support during his PhD. Y. Zhang would like to acknowledge the European Commission's Marie Skłodowska-Curie Actions (MSCA), through the Marie Skłodowska-Curie Individual Fellowships (IF-EF) (H2020-MSCA-IF-2015-EF-703950-HEAPPs) under Horizon 2020. C. R. Bowen would like to acknowledge funding from the European Research Council under the European Union's Seventh Framework Programme (FP/2007-2013)/ERC Grant Agreement no. 320963 on Novel Energy Materials, Engineering Science and Integrated Systems (NEMESIS).

**References**

[1] F. K. Shaikh, S. Zeadally, Energy harvesting in wireless sensor networks: A comprehensive review, Renew. Sustain. Energy Rev. 55 (2016) 1041–1054.

[2] C. R. Bowen, H. A. Kim, P. M. Weaver, S. Dunn, Piezoelectric and ferroelectric materials and structures for energy harvesting applications, Energy Environ. Sci. 7 (2013) 25–44.

[3] C. R. Bowen, J. Taylor, E. LeBoulbar, D. Zabek, A. Chauhan, R. Vaish, Pyroelectric materials and devices for energy harvesting applications, Energy Environ. Sci. 7 (2014) 3836–3856.

[4] R. A. Islam, S. Priya, Realization of high-energy density polycrystalline piezoelectric ceramics, Appl. Phys. Lett. 88 (2006) 032903.

[5] C. R. Bowen, J. Taylor, E. Le Boulbar, D. Zabek, V. Y. Topolov, A modified figure of merit for pyroelectric energy harvesting, Mater. Lett. 138 (2015) 243–246.

[6] J. I. Roscow, Y. Zhang, J. Taylor, C. R. Bowen, Porous ferroelectrics for energy harvesting applications, Eur. Phys. J. Spec. Top. 224 (2015) 2949–2966.

[7] Y. Zhang, M. Xie, J. Roscow, K. Zhou, Y. Bao, D. Zhang, C. Bowen, Enhanced pyroelectric and piezoelectric properties of PZT with aligned porosity for energy harvesting applications, J. Mater. Chem. A 5 (2017) 6569– 6580.

[8] G. H. Haertling, Ferroelectric Ceramics: History and Technology, J. Am. Ceram. Soc. 82 (1999) 797–818.

[9] R. W. C. Lewis, A. C. E. Dent, R. Stevens, C. R. Bowen, Microstructural modelling of the polarization and properties of porous ferroelectrics, Smart Mater. Struct. 20 (2011) 085002.

[10] J. I. Roscow, R. W. C. Lewis, J. Taylor, C. R. Bowen, Modelling and fabrication of porous sandwich layer barium titanate with improved piezoelectric energy harvesting figures of merit, Acta Mater. 128 (2017) 207–217.

[11] E. Mercadelli, A. Sanson, C. Galassi, Chapter 6: Porous piezoelectric ceramics, in: Piezoelectric Ceramics, InTech Open, 2010.

[12] R. E. Newnham, D. P. Skinner, L. E. Cross, Connectivity and piezoelectric-pyroelectric composites, Mater. Res. Bull. 13 (1978) 525–536. [12] S. Marselli, V. Pavia, C. Galassi, E. Roncari, F. Cranciun, G. Guidarelli, Porous piezoelectric ceramic hydrophone, J. Acoust. Soc. Am. 106 (1999) 733–738.

[13] S. Marselli, V. Pavia, C. Galassi, E. Roncari, F. Cranciun, G. Guidarelli, Porous piezoelectric ceramic hydrophone, J. Acoust. Soc. Am. 106 (1999) 733–738.

[14] H. Kara, R. Ramesh, R. Stevens, C. R. Bowen, Porous PZT ceramics for receiving transducers, IEEE Trans. Ultrason., Ferroelect., Freq. Control 50 (2003) 289–296.

[15] C. R. Bowen, A. Perry, A. C. F. Lewis, H. Kara, Processing and properties of porous piezoelectric materials with high hydrostatic figures of merit, J. Eur. Ceram. Soc. 24 (2004) 541–545.

[16] T. Zeng, X. Dong, S. Chen, H. Yang, Processing and piezoelectric properties of porous PZT ceramics, Ceram. Int. 33 (2007) 395–399.

[17] A.-K. Yang, C.-A. Wang, R. Guo, Y. Huang, C.-W. Nan, Porous PZT Ceramics with High Hydrostatic Figure of Merit and Low Acoustic Impedance by TBA-Based Gel-Casting Process, J. Am. Ceram. Soc. 93 (2010) 1427–1431.

[18] A. Navarro, R. W. Whatmore, J. R. Alcock, Preparation of functionally graded PZT ceramics using tape casting, J. Electroceram. 13 (2004) 413–415.

[19] C. P. Shaw, R. W. Whatmore, J. R. Alcock, Porous, Functionally Gradient Pyroelectric Materials, J. Am. Ceram. Soc. 90 (2007) 137–142.

[20] J. I. Roscow, J. Taylor, C. R. Bowen, Manufacture and characterization of porous ferroelectrics for piezoelectric energy harvesting applications, Ferroelectr. 498 (2016) 40–46.

[21] Y. Zhang, Y. Bao, D. Zhang, C. R. Bowen, Porous PZT Ceramics with Aligned Pore Channels for Energy Harvesting Applications, J. Am. Ceram. Soc. 98 (2015) 2980–2983.

[22] R. Guo, C. A. Wang, A. Yang, Effects of pore size and orientation on dielectric and piezoelectric properties of 1-3 type porous PZT ceramics, J. Eur. Ceram. Soc. 31 (2011) 605–609.

[23] S. H. Lee, S. H. Jun, H. E. Kim, Y. H. Koh, Piezoelectric properties of PZT-based ceramic with highly aligned pores, J. Am. Ceram. Soc. 91 (2008) 1912–1915.

[24] S.-H. Lee, S.-H. Jun, H.-E. Kim, Y.-H. Koh, Fabrication of porous PZT-PZN piezoelectric ceramics with high hydrostatic figure of merits using camphene-based freeze casting, J. Am. Ceram. Soc. 90 (2007) 2807–2813.

[25] W. Liu, N. Li, Y. Wang, H. Xu, J. Wang, J. Yang, Preparation and properties of 3-1 type PZT ceramics by a self-organization method, J. Eur. Ceram. Soc. 35 (2015) 3467–3474.

[26] W. Liu, L. Lv, Y. Li, Y. Wang, J. Wang, C. Xue, Y. Dong, J. Yang, Effects of slurry composition on the properties of 3-1 type porous PZT ceramics prepared by ionotropic gelation, Ceram. Int. 43 (2017) 6542–6547.

[27] I. S. Grant, W. R. Phillips, Chapter 2: Dielectrics, in: Electromagnetism, Wiley, Chichester, UK, 1990.

[28] L. Padurariu, L. P. Curecheriu, L. Mitoseriu, Nonlinear dielectric properties of paraelectric-dielectric composites described by a 3D Finite Element Method based on Landau-Devonshire theory, Acta Mater. 103 (2016) 724–734.

[29] F. Gheorghiu, L. Padurariu, M. Airimioaei, L. Curecheriu, C. Ciomaga, C. Padurariu, C. Galassi, L. Mitoseriu, Porosity-Dependent Properties of Nb-Doped Pb(Zr,Ti)O3 Ceramics, J. Am. Ceram. Soc. 100 (2016) 647–658.

[30] R. Khachaturyan, S. Zhukov, J. Schultheiß, C. Galassi, C. Reimuth, J. Koruza, H. von Seggern, Y. A. Genenko, Polarization-switching dynamics in bulk ferroelectrics with isometric and oriented anisometric pores, J. Phys. D: App. Phys. 50 (2017) 045303.

[31] C. Padurariu, L. Padurariu, L. Curecheriu, C. Ciomaga, N. Horchidan, C. Galassi, L. Mitoseriu, Role of the pore interconnectivity on the dielectric, switching and tunability properties of PZTN ceramics, Ceram. Int. 43 (2017) 5767–5773.

[32] T. Zeng, X. Dong, C. Mao, Z. Zhou, H. Yang, Effects of pore shape and porosity on the properties of porous PZT 95/5 ceramics, J. Eur. Ceram. Soc. 27 (2007) 2025–2029.

[33] H. L. Zhang, J.-F. Li, B.-P. Zhang, Microstructure and electrical properties of porous PZT ceramics derived from different pore-forming agents, Acta Mater. 55 (2007) 171–181.

[34] A. P. Roberts, E. J. Garboczi, Elastic properties of model porous ceramics, J. Am. Ceram. Soc. 83 (2000) 3041–3048.

[35] S. Deville, E. Saiz, R. K. Nalla, A. P. Tomsia, Freezing as a path to build complex composites, Science 311 (2006) 515–518.

[36] J. Seuba, S. Deville, C. Guizard, A. J. Stevenson, Mechanical properties and failure behavior of unidirectional porous ceramics, Sci. Rep. 6 (2016) 24326.

[37] D. Berlincourt, H. A. Krueger, C. Near, Properties of Morgan electro ceramic ceramics, Technical Publication TP-226, Morgan Electroceramics (1999) 1–12.

[38] T. Hang, J. Glaum, Y. A. Genenko, T. Phung, M. Hoffman, Investigation of partial discharge in piezoelectric ceramics, Acta Mater. 102 (2016) 284–291.

[39] S. Deville, E. Saiz, A. P. Tomsia, Ice-templated porous alumina structures, Acta Mater. 55 (2007) 1965–1974.

[40] S. Deville, Freezing Colloids: Observations, Principles, Control, and Use, Springer, 2017.

[41] S. Deville, Freeze-Casting of Porous Ceramics: A Review of Current Achievements and Issues, Adv. Eng. Mater. 10 (2008) 155–169.

[42] T. Waschkies, R. Oberacker, M. J. Hoffmann, Control of lamellae spacing during freeze casting of ceramics using double-side cooling as a novel processing route, J. Am. Ceram. Soc. 92 (2009) 79–84.

[43] S. Deville, E. Maire, G. Bernard-Granger, A. Lasalle, A. Bogner, C. Gauthier, J. Leloup, C. Guizard, Metastable and unstable cellular solidification of colloidal suspensions, Nat. Mater. 8 (12) (2009) 966–972.

[44] N. O. Shanti, K. Araki, J. W. Halloran, Particle redistribution during dendritic solidification of particle suspensions, J. Am. Ceram. Soc. 89 (8) (2006) 2444–2447.

[45] X.-Y. Zhang, Y. Zhang, D. Zhang, Effect of particle size on the lamellar pore microstructure of porous $Al_2O_3$ ceramics fabricated by the unidirectional freezing, App. Mech. Mater. 184-185 (2012) 818–825.

[46] C. R. Bowen, A. C. Dent, R. Stevens, M. G. Cain, A. Avent, A new method to determine the un-poled elastic properties of ferroelectric materials, Sci. Tech. Adv. Mater. (2017) 253–263.

[47] H. R. Gallantree, Piezoelectric ceramic/polymer composites, British Ceramics Proceedings 41 (1989) 161.

[48] BSI, Advanced technical ceramics – Monolithic ceramics – General and textural properties, Part 2: Determination of density and porosity (1993).

[49] A. Goncharenko, V. Lozovski, E. Venger, Lichtenecker's equation: applicability and limitations, Optics Communications (January) (2000) 19–32.

[50] G. Arlt, D. Hennings, G. D. With, Dielectric properties of finegrained barium titanate ceramics Dielectric properties of fine-grained barium titanate ceramics, J. Appl. Phys. 58 (1985) 1619–1625.

[51] A. Bell, Grain Size Effects in Barium Titanate - Revisited, IEEE (1991) 14–17.

[52] Y. Tan, J. Zhang, Y. Wu, C. Wang, V. Koval, B. Shi, H. Ye, R. McKinnon, G. Viola, H. Yan, Unfolding grain size effects in barium titanate ferroelectric ceramics, Sci. Rep. 5 (2015) 9953.

[53] D. Berlincourt, H. Jaffe, Elastic and piezoelectric coefficients of single crystal Barium Titanate, Phys. Rev. 111 (1958) 143–148.

[54] S. Wada, K. Takeda, T. Tsurumi, T. Kimura, Preparation of [110] grain oriented barium titanate ceramics by templated grain growth method and their piezoelectric properties, Japan. J. App. Phys. 46 (2007) 7039–7043.

[55] Y. Huan, X. Wang, J. Fang, L. Li, Grain size effects on piezoelectric properties and domain structure of $BaTiO_3$ ceramics prepared by two-step sintering, J. Am. Ceram. Soc. 96 (2013) 3369–3371.

[56] S. Priya, Criterion for Material Selection in Design of Bulk Piezoelectric Energy Harvesters, IEEE Trans. Ultrason., Ferroelect., Freq. Control 57 (2010) 2610–2612.